# Data modelling recipes for SARS-CoV-2 wastewater-based epidemiology

Wolfgang Rauch*^, Hannes Schenk*, Heribert Insam**, Rudolf Markt** and Norbert Kreuzinger***

* Unit of Environmental Engineering, Department of Infrastructure, University of Innsbruck, Technikerstrasse 13, 6020 Innsbruck, Austria;
** Department of Microbiology, University of Innsbruck, Austria
*** Institute for Water Quality and Resource Management, Technische Universität Wien, Austria
^ Corresponding author	(E-mail: wolfgang.rauch@uibk.ac.at)

**Abstract:**

Wastewater based epidemiology is recognized as one of the monitoring pillars, providing essential information for pandemic management. Central in the methodology are data modelling concepts for both communicating the monitoring results but also for analysis of the signal. It is due to the fast development of the field that a range of modelling concepts are used but without a coherent framework. This paper provides for such a framework, focusing on robust and simple concepts readily applicable, rather than applying latest findings from e.g., machine learning. It is demonstrated that data preprocessing, most important normalization by means of biomarkers and equal temporal spacing of the scattered data, is crucial. In terms of the latter, downsampling to a weekly spaced series is sufficient. Also, data smoothing turned out to be essential, not only for communication of the signal dynamics but likewise for regressions, nowcasting and forecasting. Correlation of the signal with epidemic indicators require multivariate regression as the signal alone cannot explain the dynamics but simple linear regression proofed to be a suitable tool for compensation. It was also demonstrated that short term prediction (7 days) is accurate with simple models (exponential smoothing or autoregressive models) but forecast accuracy deteriorates fast for longer periods.

**Highlights**

- Data preprocessing is an essential step in a modeling framework for wastewater-based epidemiology
- Equal spacing of the scatted data is essential but downsampling is sufficient
- Data smoothing is key for regression and forecast
- For long term WBE data series, multivariate regression is essential but linear regression is sufficient
- Short term signal prediction (7days) is reasonable accurate

# 1 Introduction

Since early in the appearance of the SARS-CoV-2 pandemic in the beginning of 2020, wastewater-based epidemiology (WBE) was investigated as possible tool for early warning of Covid-19 outbreaks (Medema et al., 2020). It has been demonstrated that SARS-CoV-2 ribonucleic acid (RNA) is found in wastewater at the influent of treatment plants with the genetic material being introduced into the sewer systems by viral shedding of Covid-19 patients (e.g., Wölfel et al., 2020). Based on those references, WBE has been applied in pandemic management early on, see e.g., Ahmed et al., 2020; Medema et al., 2020. In the meantime, the information from WBE is displayed in (national) dashboards (Naughton et al., 2021) and used for subsequent assessment of pandemic development (e.g., Robotto et al., 2022; Lastra et al., 2022).

Despite the successful application, there is still remarkable uncertainty in analysis and prediction of the viral signal. Key problems are the inherent errors and uncertainties in the raw virus signal as measured at the treatment plant. Other issues of concern are the link of the virus signal to epidemic indicators such as incidence values and the prediction of pandemic dynamics, which proved to be not straightforward but highly dynamic during different phases.

While a stringent methodology regarding data modelling cannot explain the inherent uncertainties of WBE, it avoids to introduce new and additional error sources. Accordingly, this paper aims to establish a modelling framework for WBE signal analysis, interpretation and prediction. Key emphasis is given to test and demonstrate mathematical methods of various complexity.

As data source we will use the case study Vienna, Austria where genetic material in the wastewater have been sampled at the main wastewater treatment plant for appr. 18 months. The timeseries thus encompasses the main phases of the Covid-19 pandemic. However, no emphasis will be put here on the details and uncertainties arising from sampling and laboratory processes (PCR test deriving genetic material). This is not to disregard the issue but the presented investigation starts with the timeline of the virus signal derived therefrom.

## 2  Materials and Methods

### 2.1  Wastewater based epidemiology for SARS-Cov-2

WBE aims to derive conclusions on substances occurring in the watershed of a sewer system by sampling – usually at the influent of a treatment plant. The concept has been originally developed for the detection of drug use (Daughton, 2001) but is in the meantime likewise applied for spreading of diseases (Sim and Kasprzyk-Hordern, 2020). Adapting the basic formulation for Sars-Cov-2, the timeseries from WBE is usually expressed as (measured) virus load related to the population drained with the sewer system:

$$L_{virus} = \frac{c_{virus} * Q}{P}$$

where $L_{virus}$ is gene copies/P/d; $Q$ = flow volume in L/d; $c_{virus}$ = virus concentration in the sample in gene copies/L and $P$ = number of persons in the watershed. While not directly comparable to an epidemic indicator, the virus load serves as quantitative information on pandemic dynamics in the watershed.

Assuming that each infected person is shedding a certain load of genetic material per time ($L_{shed}$ in Gene copies/$P_{inf}$/d) into the sewer system as well as neglecting uncertainties, distortions from variations of shedding behaviours and losses in the transport phase we get the relation:

$$L_{shed} = \frac{c_{virus}*Q}{P_{inf}} = \frac{c_{virus}*Q}{P*f_{inf}} = \frac{L_{virus}}{f_{inf}}$$

with $L_{shed}$ in gene copies/$P_{inf}$/d; $P_{inf}$ = infected persons in the watershed and $f_{inf} = P_{inf}/P$ = fraction of infected persons. Note that $f_{inf}$ is actually expressing prevalence (defined as ratio of infected persons to total population) – one of the key pandemic parameters.

### 2.2  Wastewater surveillance – the Vienna case study

Like many other countries, Austria has initiated wastewater-based epidemiology already early in the pandemic i.e., in late spring 2020. While the program started initially as research program, results from initially bi-weekly surveillance (three-weekly since November 2020) have subsequently been taken up in pandemic management. For this paper we will utilize the timeseries of SARS-Cov-2 measurements of one specific treatment plant as case study, i.e., of Vienna (1.9 Mio inhabitants) as the capital city of Austria. The length of the time series is app. 18 months, including three pandemic waves, several lockdown periods and the bulk of the vaccination campaign.

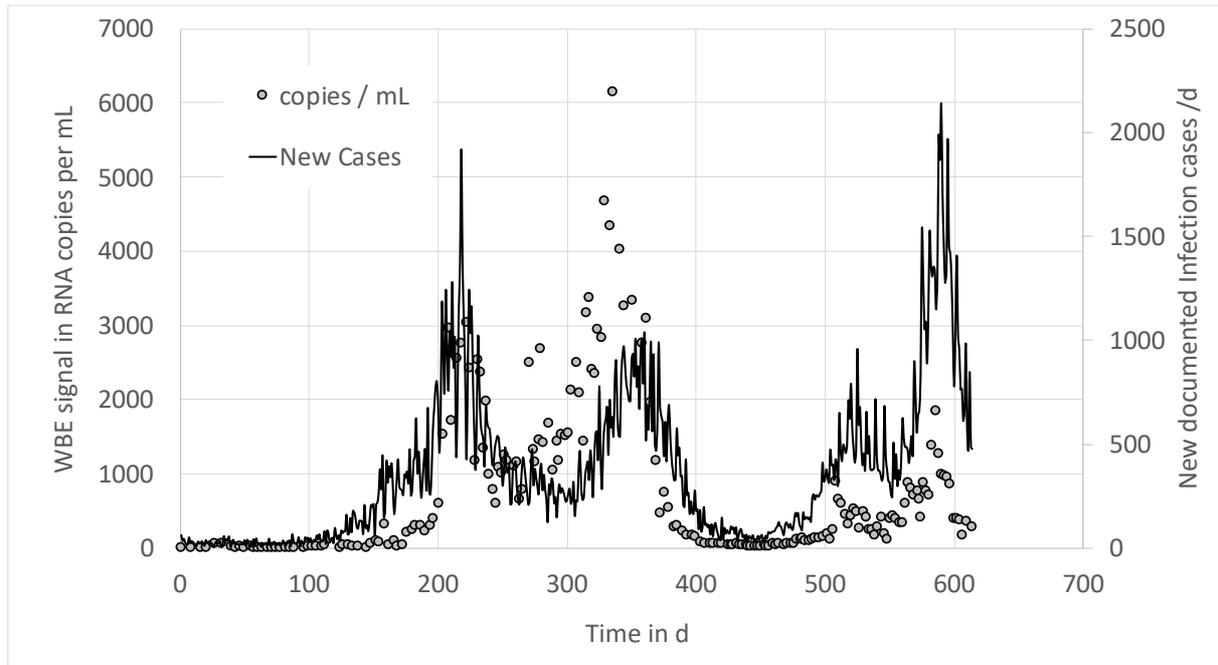

*Figure 1: Timeline of SARS-Cov-2 measurement (raw data in copies/ml) and daily documented new infections.*

### 2.3 Modeling framework

Key aspect in modelling epidemic dynamics is the understanding of the inherent information and statical properties of the signal. The virus signal – derived from WBE – is a proxy for the overall prevalence (with prevalence here defined as the fraction of infected persons in the total population). But while prevalence is a valuable information in pandemic management, the quantitative determination of the value requires huge monitoring effort (Rippinger et al., 2021) and is thus rarely done. Incidence, on the other hand, is the common pandemic indicator based on documented infection cases by human/individual tests (see Figure). The value is usually given as sum of documented infections in a 7-day period per 100 000 Persons, thus applying a gliding mean smoothing filter to the time series. However, note that incidence deviates from the actual infection status (i.e., prevalence) as the huge number of undocumented cases are not accounted for. Moreover, the incidence values are subject to test numbers and strategies and thus contain a significant uncertainty both in assessment and also in interpretation.

From the above we can derive two typical tasks for time series analysis in relation to WBE, that is first correlating the signal to pandemic indicators and second forecasting the pandemic:

- Correlating the signal to pandemic indicators relates to the standard statistical modelling problem of regression analysis, i.e., estimating the relation of the dependent variable y (e.g., incidence values) to the independent variable x (WBE signal). Applying a linear relation and fitting the parameters by the least-squares method is denoted univariate linear regression. Adding more independent variables $x_i$ (e.g., the time series of daily individual tests) leads to multivariate regression. However, the more independent variables are added the more complex the interpretation of results. Next to the linear estimation there are a huge number of other methods available – most of them from the field of machine learning. Aberi et al., (2021) compare 8 different methods in the context of WBE regression and find $3^{rd}$ order polynomial regression as outperforming.
- Predicting epidemic development is key in pandemic management, in order to plan strategic measures such as restrictions and public interventions. WBE can hardly give a full picture of infection trends under multiple influences but the forecast of the signal is adding to the general trend prediction. Hence, for prediction modelling we concern ourself with the virus signal itself. As for the above-mentioned regression problem, there are a plethora of methods and algorithms available. Parmezan et al. (2019) categorize the models into parametric ones, i.e., exponential smoothing, autoregression (AR) and moving average (MA) and non-parametric ones, i.e., machine learning approaches such as artificial neural networks and support vector machine.

### 2.4   Model performance metrics

Time series analysis and data modeling as discussed require metrics and methods for performance assessment. Basic issues concern the assessment of the fit of a data model, similarity of series expressed as correlation, cross-validation and model comparison. The metrics used and their background are described in detail in Appendix A.

# 3 Results and Discussion

## 3.1 Data Preprocessing

It is obvious from the complex processes and degrees of freedom associated with WBE that the raw signal contains both insufficient information and inherent variation, thus rendering it unsuited for further analysis as is. In the following, we propose a 4-step procedure for data pre-processing – adapted from Rauch et al, 2021, Markt et al., 2021 and Arabzadeh et al, 2021:

### 3.1.1 Normalization

For subsequent analysis with respect to epidemic indicators the accurate estimation of the population in the catchment is essential. While the relative fluctuation of persons in the watershed is usually small in large cities – despite commuters and tourists - , the population is changing widely in smaller settlements and touristic regions. The use of population biomarkers is thus standard practice to account for such variations (Been et al., 2014). While there is a range of biomarkers possible to compensate for dilutions in case of stormwater events or fluctuations in the shedding population (Choi et al., 2018), the choice is usually limited to routinely monitored water quality parameters such as COD, $NH_4$-N and $N_{tot}$ applying a standard load of 120g COD, 8g $NH_4$-N or 11 g N respectively per PE (person equivalent) per day for estimation. Rauch et al, 2021 and Arabzadeh et al, 2021 advocate the use of $NH_4$-N as more reliable parameter in this context and derive the normalized signal as

$$L_{virus} = \frac{c_{virus} * f_{NH4}}{c_{NH4}} \text{ in gene copies/PE/day}$$

where $c_{NH4}$ is the concentration of $NH_4$-N in gN/L and $f_{NH4}$ = specific $NH_4$-N load in gN/PE/d. (Use of COD and $N_{tot}$ is similar).

However, normalisation is hardly ever straightforward in a long-term monitoring campaign, as specific data are occasionally missing, plain wrong or heavily influenced by sewer flushing, differences in industrial wastewater share and alike (Amoah et al., 2022). One obvious recipe is – for those instances - to either switch automatically from one water quality parameter to another (e.g., from $NH_4$-N to COD) or to use the mean of the (other) available parameters instead. But before applying such an algorithm note that the above mention standard load values are assuming standard wastewater composition and should (in theory) result in similar population numbers when used for a specific sample, with

$$PE = \frac{c_{bm} * Q}{f_{bm}}$$

where, $c_{bm}$ is the concentration of biomarker in g/L and $f_{bm}$ = specific biomarker load in g/PE/d. Thus, prior to interchangeable use of the standard loads it is advised to check if that relation holds – and in case not – derive better load estimates. E.g., for the timeseries of Vienna the deviation of the mean of each parameter against the averaged mean of all 3 quality parameters is COD = 1.10; NH$_4$-N = 0.939 and N$_{tot}$ = 0.962 (when the standard load estimates are applied as indicated above). For switching parameters in the normalization process, such factors are to be considered.

### 3.1.2 Outlier detection

Outliers in the signal timeline (that clearly distort the subsequent information and need to be removed) can derive from various sources and not all of those errors can be tracked by means of a coherent strategy. A manual inspection of the data thus remains an essential step. One potential source of erratic data stems from extreme discharge conditions as given by heavy rain events in combined sewer system. In a pragmatic approach Rauch et al, 2021 and Markt et al.,2021 concluded that samples taken while the inflow is higher than the 90percentile of the recorded inflow data are potentially erroneous and to be treated as outliers. The shortcoming is that the determination of the threshold value (90percentile) requires at least one full year of flow data prior to the SARS-Cov-2 monitoring.

Intuitively one would apply a standard outlier detection method – such as a simple box plot (Tukey, 1970) - consecutively for the normalized signal as derived from above normalization. Once a data point is labeled as outlier (i.e., being outside the 1.5 interquartile range), alternatives would apply, such as replacing the signal value with the mean of neighboring values or appropriate quartiles. Alas, this simple approach does not work in an algorithmic setting: in the rising limb of a pandemic wave, each signal is higher as the previous one – thus all values would be labeled as outliers.

However, the strategy can be used for the biomarker value instead for the signal itself. I.e., if a ($NH_4$-N) measurement is identified as outlier, the use of other biomarkers (e.g., COD, $N_{tot}$) is advocated and - if that fails too - plausible values are to be used instead. In the latter case e.g., the 95th percentile concentration from the data series could be used as surrogate value, if the concentration values are excessively high. For the Vienna case study, the 95th percentiles would be 859 mg COD/l, 63.1 mg N/l and 45.3 mg $NH_4$-N/l. But note, the above only works for biomarker measurements not for outlier in the virus data itself.

### 3.1.3 Resampling and Interpolation of scattered data

Most standard analysis methods require equally spaced samples in the time series. In WBE on the other hand samples are taken in different frequency (usually between daily to weekly) according to pandemic necessity, thus resulting in irregular data intervals (a so-called scattered series). The reality of irregular sampling requires to apply resampling routines from an equally spaced data grid (Benedict et al., 2000; Broersen and Bos, 2006, Rehfeld et al., 2011). Two simple approaches can be applied in the context of WBE:

- Downsampling is to condense the data into larger time segments as being sampled, here into a weekly series. An intuitive yet robust approach is applying block-wise averaging, i.e., to average all measurements in a time slot (here one week) to a mean value. The shortcoming of this approach is the loss of information due to the simple averaging, but on the other hand, the averaging can easily be done manually (even if it is tedious). The application of more complex methods such as finite impulse response (FIR) filters is not necessary as the WBE signal lacks higher frequencies that could cause aliasing.
- The second approach is to apply a common interpolation procedure in order to derive an equidistance set of values (here WBE signals). Note that interpolation can in fact be used both for upsampling (i.e., deriving daily values) and downsampling (weekly signals as above). Lepot et al., 2017, give an exhaustive overview on interpolation methods. A basic approach is linear interpolation which is easily done as being implemented as function in spreadsheets. Given a high frequency of measurements (as in the Vienna series) the method exhibits nearly identical results as applying a more complex method such as the Shepard filter (Shepard, 1969) for interpolation of the scattered data (see Figure 2).

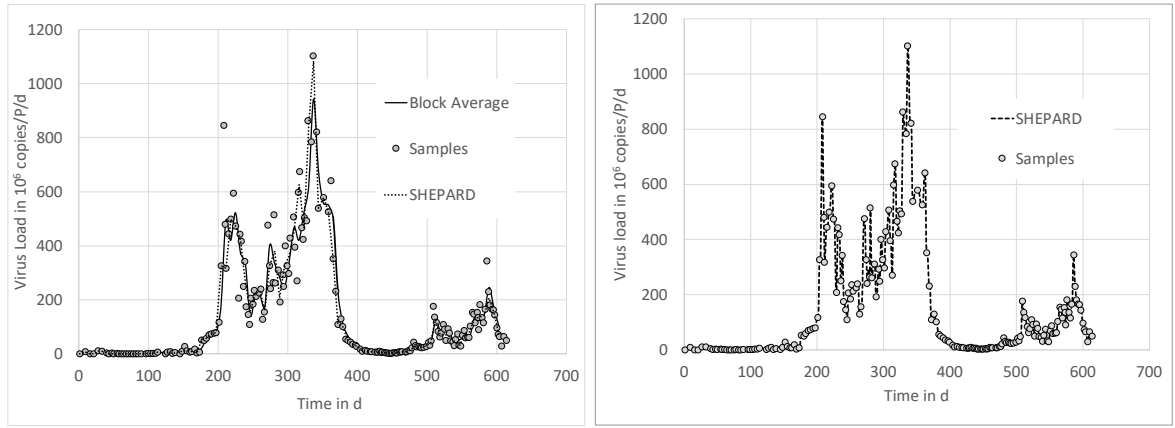

*Figure 2: Left: Downsampling (block-wise averaging) and Interpolation (Shepard Filter) to derive an equally spaced weekly time series for the Vienna case study. Right: Upsampling by means of Interpolation (Shepard Filter) towards a daily series.*

We will discuss the influence of the grid spacing (daily versus weekly series) to further data analysis (regression and prediction) below. At this point it is to be noted that both methods (Downsampling and Interpolation) imply some loss of information but are not to be confused with data filtering (smoothing) methods as the aim is different.

### 3.1.4 Data smoothing

Even after normalization, outlier detection and grid spacing the resulting time series contains a significant amount of noise, thus blurring the actual information. In order to differentiate noise from the actual information in the signal, data filtering (smoothing) is frequently applied in science and engineering (see Arabzadeh et al. 2021 for a detailed investigation in relation to WBE). Regression and prediction methods inherently cope with noise but there are two issues that make data smoothing necessary: First, for communicating the result of WBE to pandemic management and public, as the display of raw signal adds to confusion. And second, the correlation of the signal to pandemic indicators that are filtered in itself as e.g., incidence values, requires likewise to smooth the signal prior to the regression (Arabzadeh et al. 2021).

The smoothing problem is specified for a series of n observations as:

$$y_i = \beta(x_i) + \varepsilon_i, \quad i = 1, \dots, n$$

where y are the actual observations, x is the vector of covariates, ε the random error with zero mean ($\mu_{(\varepsilon)} = 0$) and β the sought smoothing function. This is in fact standard regression with the independent values x being derived from the observations y.

For data smoothing two different types of methods need to be distinguished: First, methods that require an equally spaced series (as discussed above) and second methods that can cope with irregular spaced datasets and thus provide both interpolation and smoothing features. An example of the first group is the simple moving average (SMA) while locally estimated scatterplot smoothing (LOESS) is a standard procedure for the latter (Cleveland and Devlin, 1988). To emphasize again, when applying LOESS (and alike) the resampling of the scattered dataset is not necessary.

Smoothing methods can be further classified in parametric and non-parametric ones, where for the first type the performance of the smoothing - i.e., the shape of the fitted curve - is a function of one or more parameter values. SMA is the most popular example of a parametric smoother which is calculated centralized for the estimation $y_i$ as:

$$y_i = \frac{1}{k}\sum_{j=i-(k-1)/2}^{i+(k-1)/2} x_i$$

with x = input signal, y = estimation or output signal and k = number of points in average. The parameter in SMA is k (also denoted as window size), with the smoothing effect increasing with increasing size of k. Similar LOESS is a parametric method with q specifying the fraction of the data used in a local estimate ($k_L = q*n$ is the number of neighboring data points taken into account).

Non-parametric smoothing methods have the benefit that parameter values are estimated inherently, thus providing optimal curve fitting from the point of the estimator. Typical examples are Friedman's super smoother and Generalized additive models (see e.g, Arabzadeh et al. 2021).

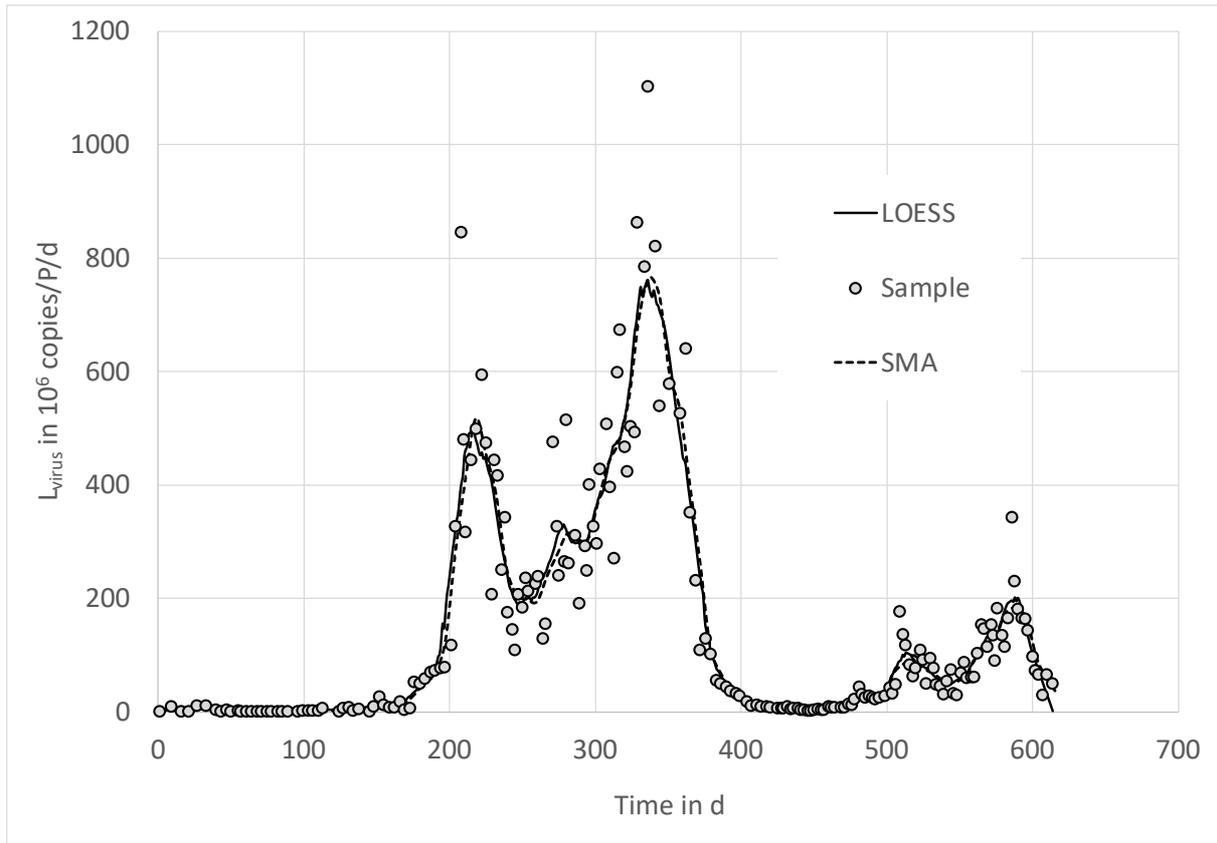

*Figure 3: Smoothing of the Vienna time series. Comparison of SMA (weekly) and LOESS (daily) with measurements (Sample).*

In Figure 3 we compare 2 methods of significantly diverging complexity: The first approach is to use the block-averaged weekly data (see Figure 2) and apply a 3-point moving average filter (k=3). The method is intuitive, robust and requires the use of spreadsheets at most. The second approach is the use of LOESS for daily data points, using 11 nearest neighbors for each local estimation ($k_L$=11). Also, LOESS is not overly complex as algorithm but definitely requires coding or the use of standard statistical routines as implemented in R or Python. (Note that there is a plethora of other and more sophisticated smoothing methods available, however the result will not differ significantly – Arabzadeh et al., 2021.)

In the case study, the optimal parameter for SMA has been determined by means of cross-validation, i.e., by minimizing LOOCV (see Appendix A) for different values of k. The parameter $k_L$ for LOESS has then been determined for highest similarity and correlation with the SMA result.

*Table 1: Key parameters of the 2 smoothing methods for the Vienna case study*

| Method | Grid spacing (d) | Parameter k, $k_L$ | RMSE | MSIM | Pearson r |
|---|---|---|---|---|---|

| SMA | 7 | 3 | 38.4 | 0.940 | 0.997 |
|---|---|---|---|---|---|
| LOESS | 1 | 11 | 60.8 | | |

Two points are to be made based on the above example: First, both methods use a kernel for deriving the estimate, which is problematic for the smoothing of the latest data (end of series). If such estimate is necessary, other methods (leaning towards prediction) must be used. And second, simple averaging methods can easily do the job – when understanding both background and restrictions of the approach. Using more complex methods does not harm, but is neither necessary nor needed for more representative outcomes (see also Smith, 1997).

### 3.2 Regression

Fundamental to mathematical regressions is the choice of the data in terms of interpolation, smoothing and variants (number of independent variables). For estimation of a suitable method, we will test 3 different datasets. Type 1 is raw data without smoothing – accordingly we hereby use the daily number of documented infections (Figure 1) as dependent variable instead of incidence values (sum of new infections over a 7-day period per 100 000 P). Type 2 are consistently smoothed values in upsampled daily grid spacing and Type 3 applies the smoothed but weekly downsampled data set. For each type the following 3 independent variables are available: first, the normalized signal itself ($L_{virus}$), second, the number of daily tests and third, the dominance of the specific virus variant B.1.1.7 alpha (Carcereny et al., 2022) in the WBE signal in percent. The last information is given by whole genome sequencing of the WBE samples (Amman et al., 2022). Exemplarily Figure 4 shows the whole dataset for Type 3 (weekly downsampled).

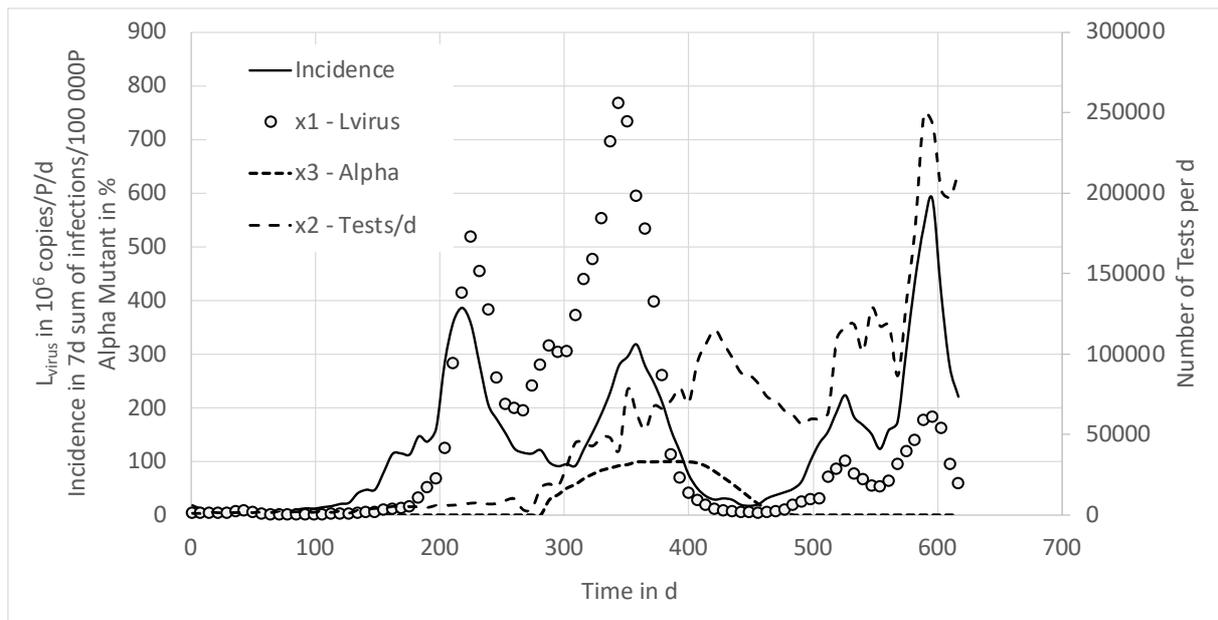

*Figure 4: Weekly spaced - smoothed dataset (Type 3) for regression analysis of the Vienna case study.*

One more step is necessary before investigation regression methods, that is the investigation of cross-correlation between the information series (WBE signal and incidence) in order to determine a possible time lag. Among others, Aberi et al., 2021 found that the incidence signal lags 2 – 7 days behind the wastewater signal which is essential to be considered. The two dominant reasons for that behavior are (Olesen et al., 2021): first, an infected person starts virus shedding before the onset of symptoms and thus before being urged to test. Second, incidence is calculated as gliding sum over 7 days which inherently delays the signal. Table 2 plots Pearson's correlation coefficient r for the 3 data sets (all are significant in terms of $p < 0.05$). But note that r is fairly constant in the range 1-8 days and thus also the time lag is a range rather than a specific value.

*Table 2: Correlation of wastewater signal with new infections (Type 1) and incidence values (Type 2/3) respectively by Pearson´s correlation coefficient r.*

| Type | Lag | r | p (<0.05) |
|---|---|---|---|
| 1 | 3 days | 0.517 | 0.079 |
| 2 | 8 days | 0.571 | 0.079 |
| 3 | 1 week | 0.558 | 0.209 |

As regression methods we compare multivariate linear regression and 3rd order polynomial regression by plotting the coefficient of determination ($R^2$). From the results summarized in Table 3 we can derive the following: first, it is due to the high variation of the raw signals (Type 1) that smoothing makes for a substantial improvement of correlation, second, spacing of the series (daily versus weekly data) does not make a substantial difference and third, polynomial regression is superior to linear regression (thus corroborating Aberi, et al., 2021).

*Table 3: Polynomial and linear regression applied to 3 different datasets (Type 1-3) of the Vienna case study*

| Typ | Lag | 3$^{rd}$ Order Polynomial | Linear Regression | | |
|---|---|---|---|---|---|
| | | $R^2$ | $R^2$ | RMSE | LOOCV |
| 1 | 3 days | 0.749 | 0.604 | 231.2 | 233.5 |
| 2 | 8 days | 0.882 | 0.755 | 62.6 | 61.2 |
| 3 | 1 (week) | 0.845 | 0.731 | 65.5 | 69.8 |

An important consideration in regression is understanding the cause-effect relation, especially in the case of multiple variants. Despite the fact that polynomial regression (and the same applies for other machine-learning based methods) gives a better correlation with the dependent variable (incidence value), the simple linear regression method is likely the superior option for analysis as being both robust and easy to interpret. We will thus use the linear regression method for further analysis of the Type 3 dataset (weekly – smoothed).

For testing robustness, we apply cross-validation and find here close values for LOOCV as compared with the RMSE value for the whole series (see Table 3). This indicats that the model predicts missing data points quite well - the deviation of the root means residuals is app. 6%. Another point is to test if the amount of information used (multiple variants) is actually necessary. A standard t-test for each of the regression coefficients (relating to $x_i$) indicates that all 3 independent variables are essential for the regression. Last, we can additionally use the 5% - 95% confidence interval for displaying the result (Figure 5).

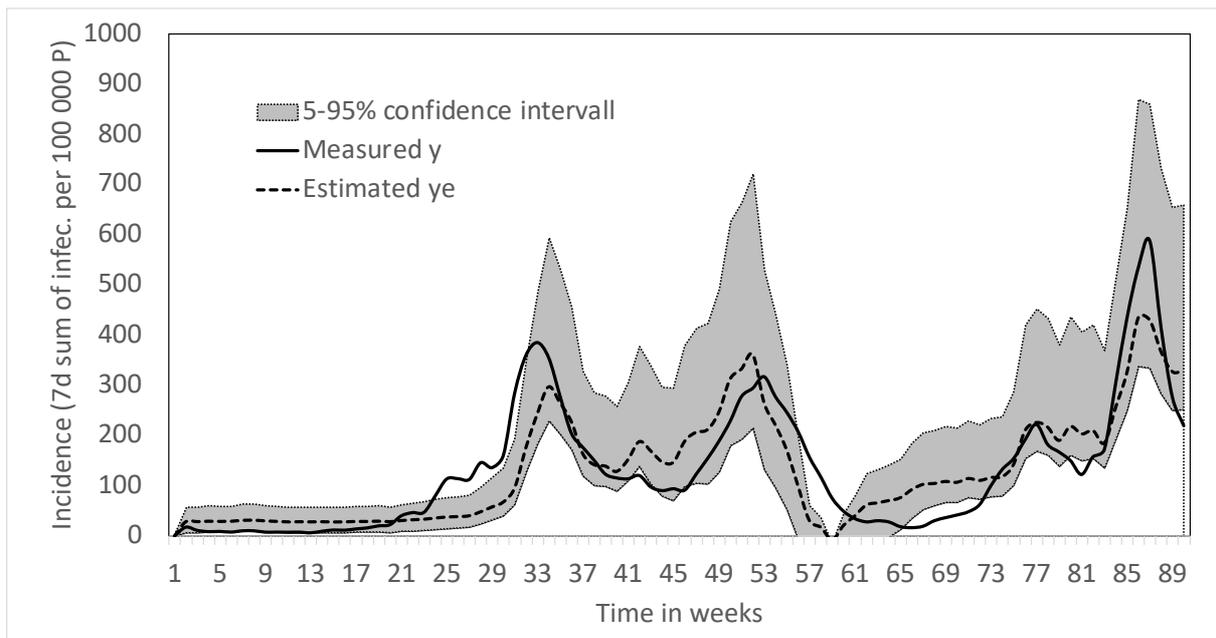

*Figure 5: Regression analysis of the Vienna case study – estimating incidence (y) with multivariate linear regression (ye).*

One last point: In this example we aimed for correlating the whole data series, i.e., including the occurrence of virus mutants and highly varying number of tests – which proofed to be difficult and requiring multivariate regression for a suitable result. However, regression is substantially easier to achieve when investigating shorter periods e.g., each of the 3 pandemic waves individually. In this case even simple univariate linear regression might easily do the job.

### 3.3 Forecast

Forecast in time series analysis is a well understood topic (De Gooijer and Hyndman, 2006) and has received already much attention in relation to the SARS-CoV-2 pandemic (see e.g., Cao and Francis, 2021). In the following we outline the procedure for the WBE signal ($L_{virus}$), thus aiming to univariate short-term forecast of appr. one to two weeks ahead. While many complex methods and models are described in the literature (Parmezan et al., 2019) in our approach we will focus on two simple yet robust linear prediction models, that is first, exponential smoothing and second, autoregressive modelling (AR). Both provide understanding of the problem at hand and are easily implemented in WBE surveillance routines.

Simple exponential smoothing (SES) computes - for a time series with N observations y - the forecast $\hat{y}$ from previous observations and forecasts as

$$\hat{y}_{N+1} = \alpha * y_N + (1-\alpha) * \hat{y}_{N-1}$$

where α (and of minor importance the starting value $\hat{y}_0$) are the parameters. Note that the above describes a one-step-ahead forecast. Forecasting over several periods (e.g., 7 days ahead in a daily time step series) is here provided by iteration, i.e., the forecasted values are taken as observations for the desired number of forecast periods (Marcellino et al., 2006).

Autoregressive modelling assumes the forecast to be linear dependent on the previous observations:

$$\hat{y}_{N+1} = c + \sum_{j=1}^{p} \phi_j y_{N-j} + \epsilon_N$$

where p is the order of the model, written as AR$_{(p)}$, ϕ are the parameters, c a constant and ε the white noise. The estimation of the parameter values is straightforward by non-linear optimization, however to determine p is not direct but e.g., based on autocorrelation of the series (Box and Jenkins, 1970). Multistep prediction is accomplished as above for SES by iteration.

Prior to the application of linear prediction models, we need to check stationarity in the data. Box and Jenkins (1970) propose the following method:

Stationarity regarding the variances (i.e., normality) is conveniently determined visually by means of a Q-Q plot (Ben and Yohai, 2004) - easily performed in any standard spreadsheet. As the Vienna data series (we investigate the both the daily and weekly smoothed series in the following) does not exhibit normality, we thus need to apply a Box-Cox transformation (Box and Cox, 1964):

$$y^{(\lambda)} = \begin{cases} \frac{y^\lambda - 1}{\lambda}, & \text{if } \lambda \neq 0 \\ \log y, & \text{if } \lambda = 0 \end{cases}$$

where λ is to be derived from maximizing the following log-likelihood function aiming for normal distribution:

$$l(\lambda) = -\frac{n}{2} ln \left[ \frac{1}{n} \sum_{i=1}^{n} \left( y_i^{(\lambda)} - \frac{1}{n} \sum_{i=1}^{n} y_i^{(\lambda)} \right)^2 + (\lambda - 1) \sum_{i=1}^{n} \ln y_i \right]$$

With λ being determined by grid search optimization as 0.117 (for both the daily and weekly series), the transformed data series are (close to) linear in the Q-Q plot (Figure 6), thus demonstrating normality.

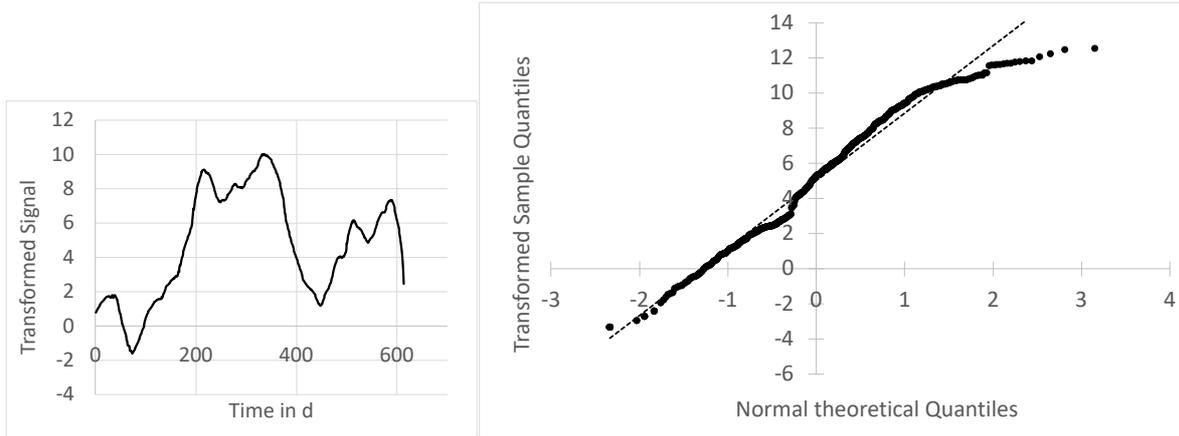

*Figure 6: Left: Box-Cox transformed $L_{virus}$ data series (daily values) and Right: Q-Q plot of the transformed series.*

Next step is to check for stationarity in the mean of the time series, i.e., if the series exhibits trend and/or seasonality. We apply here a unit root test for stationarity, namely the augmented Dickey-Fuller (ADF) test (Fuller, 1976) in the following form:

$$\Delta y_t = \alpha + \beta y_{t-1} + \gamma \Delta y_{t-1}$$

with $\Delta y_t = y_t - y_{t-1}$ being the differencing operator. The above equation relates to a bi-variate regression problem and the null-hypothesis (i.e., the series is stationary) is given if β = 0. We apply the ADF test for the transformed time series (both daily and weekly samples) and find that β is significant for the t-test and thus both series are non-stationary. In order to stabilize the mean of a time series and therefore eliminating (reducing) trend and seasonality we apply additionally differencing (see above operator). The transformed series are back-transformed after forecasting.

While the Box-Jenkins method towards stationarity is widely applied, it has also received criticism and the usefullness of the transformation is debated (Chatfield and Prothero, 1973). Hence, we use a post-sample approach (Makridakis and Hibon, 1997) and check for the optimal procedure in terms of methods (SES versus AR), data transformation (none, differencing, Box-Cox plus differencing), parameter α (SES) and order p (AR) respectively, daily or weekly timeseries and length of the forecast (7 versus 14 days). For the investigation we go through the total time series (N entries) and consecutively compute the difference between the forecast and the observation for any point i [$p_{max}$ - (N-$f_p$)] where $f_p$ = forecast periods and $p_{max}$ = maximum order of the AR model (here 10). The term post-sample relates to the forecasts, where the observations (samples) are only used for determining the error but not for training the forecast model itself. As metric we use AIC (Akaike information criterion – see Appendix A) but the display of results (see detailed data in Appendix B) is done for RSME instead - as the actual values are comparable. The result of the post-sample analysis can be summarized as follows:

- Forecast with the daily raw data series resulted in substantially worse performance as compared with the smoothed series and was thus not further analyzed.
- Data transformation is necessary but SES works best with differencing alone, while AR is optimal for BoxCox transformation plus differencing
- Optimal parameter values are dependent on spacing of series: SES α is 0.7 for daily and 0.1 for weekly series, while AR p is 3 for daily and 5 for weekly series
- AR gives better results for all situations but SES is applicable for 7-day forecasts
- Both SES and AR forecasts deteriorate significantly for 14 days prediction
- For 7-day forecasts both daily and weekly series give similar results (see Figure 7)

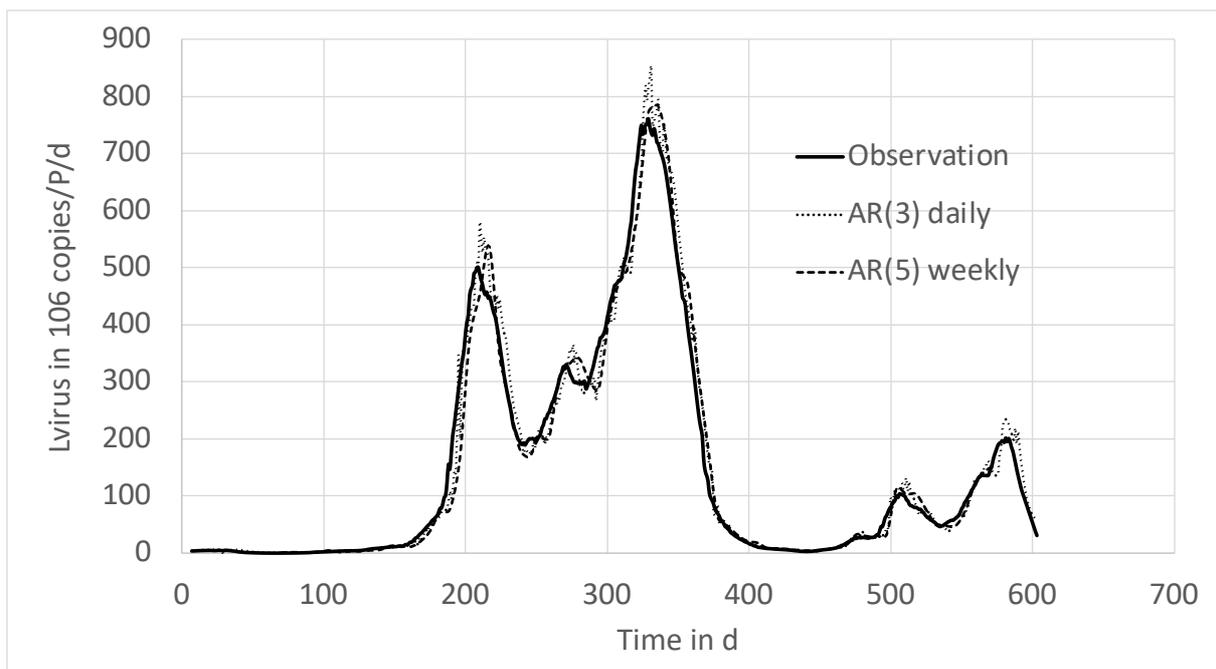

*Figure 7: Post-sample analysis of autoregressive models. The consecutive 7-day forecasts are plotted against the observation for both the daily and weekly spaced series.*

## 3.4 Recommendation

For quantitative assessment of SARS-CoV-2 wastewater surveillance data the following sequence of steps is recommended: First - and irrespective of the final aim – data preprocessing is necessary in the following order: a) normalization of the signal – including the choice of an appropriate water quality parameter as biomarker that is available as metadata, b) check for outlier values c) equal spacing of the scattered data – with preference for weekly spaced data and d) smoothing of the data series. As outlined, some smoothing methods can be applied directly to scattered data, making step c redundant. Note that smoothing (filtering) of the data series aims to capture the important patterns in the series which is not only helpful for communication but essential for subsequent data analysis.

Regression analysis aims to understand the relation of the wastewater signal with specific indicators used in pandemic management, e.g., incidence. All data series used in the statistical analysis need – prior - to be equally spaced and filtered. Next, the cross-correlation of the wastewater signal with the indicator values allows to estimate the time-lag between the data sets, which is to be taken into account. For actual regression it is recommended to apply linear regression as long as performance is sufficient in terms of the metrics AIC or coefficient of determination. Subsequent cross-validation builds confidence in the model. Last, it is important to determine the amount of additional information (additional series of independent variables) needed to explain the phenomena. Parsimony is key and the significance of each series added is to be verified (e.g., by Student's t test).

As for regression, forecasting requires an equally spaced and filtered time series of the wastewater-based signal. The use of an autoregressive model is suggested but – if a dedicated statistical software is not available – also exponential smoothing works as method. Data transformation in terms of differencing is required for both methods but AR models additional have to apply a BoxCox transformation. And last, do not forecast for more than 7 days with the simple univariate models discussed in this paper.

## 4   Conclusion

In the present study, we systematically investigated data modelling in the framework of SARS-CoV-2 wastewater-based epidemiology. Key emphasis was put on suitable recipes and methods necessary to process the signal for subsequent display in dashboards as well as correlating to pandemic indicators and for prediction. The analysis was made for the case study of an 18-month data series from Vienna, Austria, providing the WBE signal itself as well as wastewater metadata and pandemic indicator values. Key findings are as follows:

- Careful data preprocessing is essential. This implies not only normalization (by means of biomarkers) but also to provide for equal spacing of the scattered data and appropriate smoothing methods. The latter proofed to be essential for regression and forecast.
- Equal spacing of the scattered data is key but there is little to gain from detailed upsampling to e.g., daily values. Simple downsampling towards a weekly series gives nearly similar result and is substantially easier to handle.
- Correlation of the WBE signal with the pandemic indicator "incidence" required multivariate regression when using the whole time series. Albeit regression by polynomial is superior, simple linear regression is likely the better option in terms of understanding and interpretation.
- Univariate signal forecasting works well for short-term (7 day) prediction but deteriorates fast for longer time spans. Both AR and SES models can be used but AR revealed consistently better results.
- Last, standard WBE data modelling hardly ever requires the use of complex and specialized software – and most can be done in spreadsheets. As usual, it is more important to understand the underlying concepts and reasoning than applying sophisticated algorithms.

**Acknowledgements**

This study was funded by the Austrian Federal Ministry of Education, Science and Research. We would like to thank the staff of the treatment plant involved for their support and the use of the wastewater titer measurement data.

# References

Aberi, P., Arabzadeh, R., Insam, H., Markt, R., Mayr, M., Kreuzinger, N. and Rauch, W., 2021. Quest for Optimal Regression Models in SARS-CoV-2 Wastewater Based Epidemiology. International journal of environmental research and public health, 18(20), p.10778.

Ahmed, W., Angel, N., Edson, J., Bibby, K., Bivins, A., O'Brien, J.W., Choi, P.M., Kitajima, M., Simpson, S.L., Li, J. and Tscharke, B., 2020. First confirmed detection of SARS-CoV-2 in untreated wastewater in Australia: a proof of concept for the wastewater surveillance of COVID-19 in the community. Science of The Total Environment, 728, p.138764.

Amman, F., Markt, R., Endler, L., Hupfauf, S., Agerer, B., Schedl, A., Richter, L., Zechmeister, M., Bicher, M., Heiler, G. and Triska, P., 2022. National-scale surveillance of emerging SARS-CoV-2 variants in wastewater. medRxiv.

Amoah, I.D., Abunama, T., Awolusi, O.O., Pillay, L., Pillay, K., Kumari, S. and Bux, F., 2022. Effect of selected wastewater characteristics on estimation of SARS-CoV-2 viral load in wastewater. Environmental Research, 203, p.111877.

Arabzadeh, R., Gruenbacher, D.M., Insam, H., Kreuzinger, N., Markt, R. and Rauch, W., 2021. Data filtering methods for SARS-CoV-2 wastewater surveillance. Water Sci Technol (2021) 84 (6): 1324–1339.

Been, F., Rossi, L., Ort, C., Rudaz, S., Delémont, O. and Esseiva, P., 2014. Population normalization with ammonium in wastewater-based epidemiology: Application to illicit drug monitoring. Environmental science & technology, 48(14), pp.8162-8169.

Ben, M.G. and Yohai, V.J., 2004. Quantile–quantile plot for deviance residuals in the generalized linear model. Journal of Computational and Graphical Statistics, 13(1), pp.36-47.

Benedict, L.H., Nobach, H. and Tropea, C., 2000. Estimation of turbulent velocity spectra from laser Doppler data. *Measurement Science and Technology*, *11*(8), p.1089.

Box, G.E.P. and D.R. Cox. 1964. An analysis of transformations. *Journal of the Royal Statistical Society B* 26: 211-252.

Box, G.E., Jenkins, G.M., 1970. Time series analysis: forecasting and control. *Holden-Day, San Francisco.*

Broersen, P.M. and Bos, R., 2006. Estimating time-series models from irregularly spaced data. *IEEE transactions on instrumentation and measurement*, *55*(4), pp.1124-1131.

Burnham K P, Anderson D R. 2002. Model Selection and Multimodel Inference: A Practical Information-theoretic Approach, 2nd ed. Springer, New York. p. 49–89

Cao, Y. and Francis, R., 2021. On forecasting the community-level COVID-19 cases from the concentration of SARS-CoV-2 in wastewater. Science of The Total Environment, 786, p.147451.

Carcereny, A., Garcia-Pedemonte, D., Martínez-Velázquez, A., Quer, J., Garcia-Cehic, D., Gregori, J., Antón, A., Andrés, C., Pumarola, T., Chacón-Villanueva, C. and Borrego, C.M., 2022. Dynamics of SARS-CoV-2 Alpha (B. 1.1. 7) variant spread: The wastewater surveillance approach. Environmental Research, p.112720.

Cleveland, W.S. and Devlin, S.J., 1988. Locally weighted regression: an approach to regression analysis by local fitting. *Journal of the American statistical association*, *83*(403), pp.596-610.

Daughton, C.G., 2001. Illicit drugs in municipal sewage. Pharmaceuticals and Care Products in the Environment, 791, pp.348-364.

De Gooijer, J.G. and Hyndman, R.J., 2006. 25 years of time series forecasting. *International journal of forecasting*, *22*(3), pp.443-473.

Donald Shepard, A two-dimensional interpolation function for irregularly spaced data, ACM '68: Proceedings of the 1968 23rd ACM National Conference, ACM, pages 517-524, 1969.

Fuller, W. A. (1976). Introduction to Statistical Time Series. New York: John Wiley and Sons. ISBN 0-471-28715-6.

Good, P.I., 2006. *Resampling methods*. Birkhueser Boston.

Huisman, J.S., Scire, J., Caduff, L., Fernandez-Cassi, X., Ganesanandamoorthy, P., Kull, A., Scheidegger, A., Stachler, E., Boehm, A.B., Hughes, B. and Knudson, A., 2021. Wastewater-based estimation of the effective reproductive number of SARS-CoV-2. *medRxiv*.

Jones, R.H., 1984. Fitting multivariate models to unequally spaced data. In *Time series analysis of irregularly observed data* (pp. 158-188). Springer, New York, NY.

Kumar, N. and Susan, S., 2020, July. Covid-19 pandemic prediction using time series forecasting models. In *2020 11th International Conference on Computing, Communication and Networking Technologies (ICCCNT)* (pp. 1-7). IEEE.

Lastra, A., Botello, J., Pinilla, A., Urrutia, J.I., Canora, J., Sánchez, J., Fernández, P., Candel, F.J., Zapatero, A., Ortega, M. and Flores, J., 2022. SARS-CoV-2 detection in wastewater as an early warning indicator for COVID-19 pandemic. Madrid region case study. Environmental research, 203, p.111852.

Lepot, M., Aubin, J.B. and Clemens, F.H., 2017. Interpolation in time series: An introductive overview of existing methods, their performance criteria and uncertainty assessment. *Water*, *9*(10), p.796.

Makridakis, S. and Hibon, M., 1997. ARMA models and the Box–Jenkins methodology. *Journal of forecasting*, *16*(3), pp.147-163.

Marcellino, M., Stock, J.H. and Watson, M.W., 2006. A comparison of direct and iterated multistep AR methods for forecasting macroeconomic time series. *Journal of econometrics*, *135*(1-2), pp.499-526.

Markt, R., Endler, L., Amman, F., Schedl, A., Penz, T., Büchel-Marxer, M., Grünbacher, D., Mayr, M., Peer, E., Pedrazzini, M. and Rauch, W., 2021. Detection and abundance of SARS-CoV-2 in wastewater in Liechtenstein, and the estimation of prevalence and impact of the B. 1.1. 7 variant. Journal of Water and Health.

Medema, G., Heijnen, L., Elsinga, G., Italiaander, R. and Brouwer, A., 2020. Presence of SARS-Coronavirus-2 RNA in sewage and correlation with reported COVID-19 prevalence in the early stage of the epidemic in the Netherlands. *Environmental Science & Technology Letters*, *7*(7), pp.511-516.

Naughton, C.C., Roman, F.A., Alvarado, A.G.F., Tariqi, A.Q., Deeming, M.A., Bibby, K., Bivins, A., Rose, J.B., Medema, G., Ahmed, W. and Katsivelis, P., 2021. Show us the data: Global COVID-19 wastewater monitoring efforts, equity, and gaps. *medRxiv*.

Olesen, Scott W., Maxim Imakaev, and Claire Duvallet. "Making waves: Defining the lead time of wastewater-based epidemiology for COVID-19." Water Research 202 (2021): 117433.

Parmezan, A.R.S., Souza, V.M. and Batista, G.E., 2019. Evaluation of statistical and machine learning models for time series prediction: Identifying the state-of-the-art and the best conditions for the use of each model. *Information sciences*, *484*, pp.302-337.

Rauch W., Arabzadeh R., Grünbacher D., Insam H., Markt R., Scheffknecht Ch. und Kreuzinger N. (2021): Datenbehandlung in der SARS-CoV-2 Abwasserepidemiologie. Korrespondenz Abwasser – In Druck.

Rehfeld, K., Marwan, N., Heitzig, J. and Kurths, J., 2011. Comparison of correlation analysis techniques for irregularly sampled time series. *Nonlinear Processes in Geophysics*, *18*(3), pp.389-404.

Sims, N. and Kasprzyk-Hordern, B., 2020. Future perspectives of wastewater-based epidemiology: monitoring infectious disease spread and resistance to the community level. Environment international, 139, p.105689.

Rippinger, C., Bicher, M., Urach, C., Brunmeir, D., Weibrecht, N., Zauner, G., Sroczynski, G., Jahn, B., Mühlberger, N., Siebert, U. and Popper, N., 2021. Evaluation of undetected cases during the COVID-19 epidemic in Austria. BMC Infectious Diseases, 21(1), pp.1-11.

Robotto, A., Lembo, D., Quaglino, P., Brizio, E., Polato, D., Civra, A., Cusato, J. and Di Perri, G., 2022. Wastewater-based SARS-CoV-2 environmental monitoring for Piedmont, Italy. Environmental Research, 203, p.111901.

Smith, S.W., 1997. The scientist and engineer's guide to digital signal processing.

Vio, R., Strohmer, T. and Wamsteker, W., 2000. On the reconstruction of irregularly sampled time series. *Publications of the Astronomical Society of the Pacific*, *112*(767), p.74.

Wölfel, R., Corman, V. M., Guggemos, W., Seilmaier, M., Zange, S., Müller, M. A., ... & Wendtner, C. (2020). Virological assessment of hospitalized patients with COVID-2019. Nature, 581(7809), 465-469.

**Appendix A**

Time series analysis and data modeling as discussed herein requires metrics and methods for performance assessment. The most basic issue is the assessment of the fit of a data model, i.e., the comparison of the model predictions y with the measured data x for a series with n observations. The accuracy of a model is usually estimated by the metrics Root mean square error (RMSE) and – scale independent - the Coefficient of determination (R-squared):

$$RMSE = \sqrt{\frac{1}{n}\sum_{i=1}^{n}(x_i - y_i)^2} \qquad R^2 = 1 - \frac{\sum_{i=1}^{n}(x_i - y_i)^2}{\sum_{i=1}^{n}(x_i - \bar{x})^2}$$

with $\bar{x}$ denoting the sample mean, i.e., $\bar{x} = \frac{1}{n}\sum_{i=1}^{n} x_i$

Another issue that appears herein is the similarity of 2 data sets, expressed by the metric mean similarity (MSIM):

$$MSIM = \frac{1}{n}\sum_{i=1}^{n} 1 - \frac{|y_i - x_i|}{|y_i| + |x_i|}$$

The standard measure for similarity is the linear correlation between the 2 series, which is expressed as Pearson correlation coefficient r. Pearson's r is the ratio between the covariance of the 2 datasets and the product of their standard deviations, resulting in values between 1 and -1. If there is no linear association the value is 0.

$$r = \frac{\sum_{i=1}^{n}(x_i - \bar{x})(y_i - \bar{y})}{\sqrt{\sum_{i=1}^{n}(x_i - \bar{x})^2}\sqrt{\sum_{i=1}^{n}(y_i - \bar{y})^2}}$$

For a sufficient large number of samples in the time series one can assume a normal distribution of the data. We can thus express the significance of the correlation using the standard p value 0.05, that is by testing if the r value is inside the 5% - 95% interval of the normal distribution. Just to exemplify, the cross correlation of 2 data sets with n points is significant if

$$r > \frac{2*0.98}{\sqrt{n}} = \frac{1.96}{\sqrt{n}}$$

Validation of parametric models requires to apply the trained model on a new set of data in order to test the validity of the model predictions. For regression, data smoothing and prediction, cross validation is frequently performed, where the data set is repeatedly resampled – splitting the series in subsets for training and testing. A specific variant of the method is the leave one out cross validation method (LOOCV). Here n experiments are performed with a model where exactly one of the observations ($x_i$, $y_i$) is left out for later testing and all other data points (n-1) are used for training (parameter estimation). The trained model is then used to predict the left-out observation $y_i$ with $\hat{y}_i$ denoting the predicted value. Model suitability is then estimated by summing up the error for the n tests, using e.g., the RMSE metric or alike:

$$LOOCV = \sqrt{\frac{1}{n}\sum_{i=1}^{n}(y_i - \hat{y}_i)^2}$$

Cross-validation can be used twofold: first, simply to validate a model by estimating the generalization error LOOCV and second, to select among several options the model with the smallest LOOCV.

Last, we need a metric to compare models against each other. The above-mentioned metrics RMSE and $R^2$ can be used but tend to favor overparameterized models. AIC is the most common applied method to take the number of parameters into account (Burnham and Anderson, 2002). When applied to the comparison of models based on least squares fitting, one can resort to the following simplified form

$$AIC = n * log\left(\frac{1}{n}\sum_{i=1}^{n}(x_i - y_i)^2\right) + 2k$$

with k denoting the number of parameters in the model. The best model is the one that minimizes AIC. Note if comparing models with equal numbers of k the above equation is in fact equal to RSME, i.e., the use of AIC makes only sense for comparing models of different complexity in terms of parameters.

# Appendix B

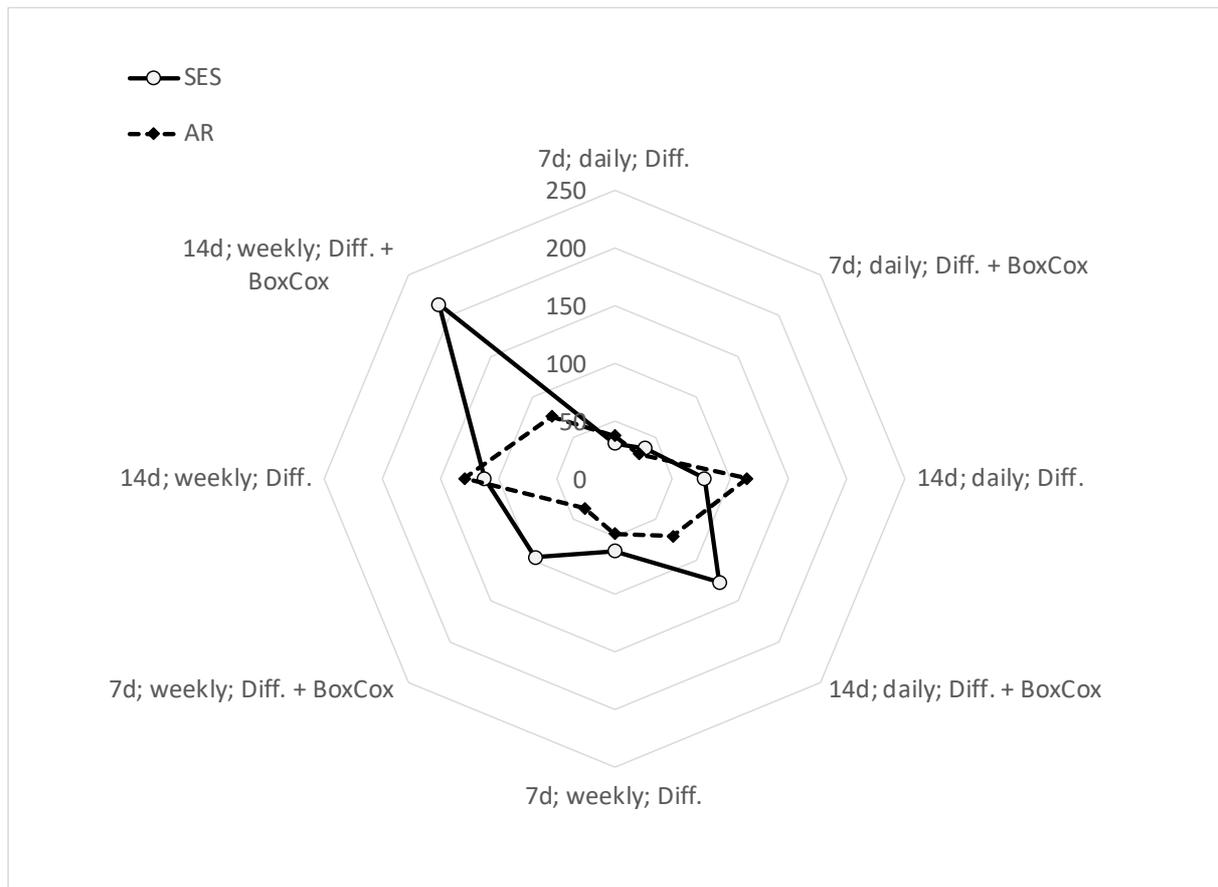

*Figure 8: RMSE result of post-sample analysis for Simple Exponential Smoothing (SES) and Autoregressive Model (AR) for daily and weekly spaced time series. Both series are smoothed. Forecast horizon 7 days and 14 days. Data transformation: Differencing and BoxCox transformation plus differencing.*